\documentclass[
 reprint,
 amsmath,amssymb,
 aps,
]{revtex4-2}
\usepackage[dvipsnames, svgnames, x11names]{xcolor}
\usepackage{graphicx}
\usepackage{dcolumn}
\usepackage{bm}

\begin{document}

\preprint{APS/123-QED}

\title{\textcolor{black}{Optimal model} for fewer-qubit CNOT gates with Rydberg atoms}
\author{Rui Li$^{1}$, Shurui Li$^{1}$, Dongmin Yu$^{1}$, Jing Qian$^{1,*}$ and Weiping Zhang$^{2,3,4}$ }

\affiliation{$^{1}$State Key Laboratory of Precision Spectroscopy, Department of Physics, School of Physics and Electronic Science, East China Normal University, Shanghai, 200062, China
}
\affiliation{$^{2}$School of Physics and Astronomy, and Tsung-Dao Lee Institute, Shanghai Jiao Tong University, Shanghai, 200240, China}
\affiliation{$^{3}$Shanghai Research Center for Quantum Science, Shanghai, 201315, China}
\affiliation{$^{4}$Collaborative Innovation Center of Extreme Optics, Shanxi University, Taiyuan, Shanxi 030006, China}



\begin{abstract}
Fewer-qubit quantum logic gate, serving as a basic unit for constructing universal multiqubit gates, has been widely applied in quantum computing and quantum information. However, traditional constructions for fewer-qubit gates often utilize a multi-pulse protocol which inevitably suffers from serious intrinsic errors during the gate execution. In this article, we report an optimal model about universal two- and three-qubit CNOT gates mediated by excitation to Rydberg states with easily-accessible van der Waals interactions. This gate depends on a global optimization to implement amplitude and phase modulated pulses via genetic algorithm, which can facilitate the gate operation with fewer optical pulses. Compared to conventional multi-pulse piecewise schemes, our gate can be realized by simultaneous excitation of atoms to the Rydberg states, saving the time for multi-pulse switching at different spatial locations. 
Our numerical simulations show that a single-pulse two(three)-qubit CNOT gate is possibly achieved with a fidelity of 99.23$\%$(90.39$\%$) for two qubits separated by 7.10 $\mu$m when the fluctuation of Rydberg interactions is excluded.
Our work is promising for achieving fast and convenient multiqubit quantum computing in the study of neutral-atom quantum technology.

\end{abstract}
\email{Corresponding author: jqian1982@gmail.com}
\pacs{}
\maketitle
\preprint{}



\section{Introduction}
Neutral Rydberg atoms with high-lying and long-range strongly-interacting properties, have long-time been considered as an ideal platform for large-scale quantum computing (see reviews {\it e.g.} refs.\cite{Saffman_2016,doi:10.1116/5.0036562,RevModPhys.82.2313}). The excitation blockade mechanism associated with Rydberg atoms can be used to prevent transitions of more than one Rydberg excitations within a whole atomic ensemble, enabling  extensive applications in the researches of multiqubit quantum gates \cite{PhysRevA.82.034307,Isenhower:11}, multiparticle entanglement \cite{PhysRevLett.100.170504,PhysRevLett.102.240502,PhysRevLett.102.170502,PhysRevA.82.062328,PhysRevLett.115.093601,PhysRevLett.123.230501} and quantum sensing \cite{RevModPhys.89.035002,2020Atomic}. So far, efforts in multiqubit quantum gates mainly focus on the construction of fewer-qubit gates which are basic logic calculation units for universal quantum computation. Yet most earlier achievements adopted the way of a piecewise pulse comprising a set of sequential $\pi$ or $2\pi$ pulses \cite{PhysRevA.88.062337,PhysRevA.82.030306,PhysRevA.92.022336}. This method relies on a rapid switching of the pulses at different spatial locations, as well as strong Rydberg blockade. Any imperfection in the blockade strength can induce a residual blockade error scaling as $\sim\Omega^2/V^2$ where $\Omega$ is the coupling strength and $V$ is the dipole-dipole interaction \cite{PhysRevA.72.022347}. 
Also these pioneering schemes are found to be vulnerable to the accumulation of decay error which increases with the number of atoms \cite{Isenhower:11}. To date, whether there exists another means for achieving fast and convenient multiqubit quantum gates other than by excitation blockade in a strong dipole-dipole blockade regime, remains an open question.

To our knowledge, adiabatic gates making use of a Rydberg dark eigenstate with fewer shaped pulses offer new opportunities, although the requirement for adiabaticity renders the gate slow and susceptible to the decay error \cite{PhysRevA.89.030301,PhysRevA.96.022321,PhysRevA.96.042306,Yu:19,PhysRevX.10.021054,PhysRevA.101.030301,PhysRevA.94.032306}. The way of shortcuts to adiabaticity can accelerate the gate operation to some extent \cite{Liao:19}. Here we adopt another acceleration approach based on pulse optimization. As an important example, we focus on the implementation of fewer-qubit CNOT gates with fewer optimally-shaped optical pulses. The importance of the CNOT gate partly stems from its ability to generate entangled states \cite{PhysRevA.82.030306}. Besides, it can basically comprise any desired gates together with single-qubit operations \cite{PhysRevA.52.3457}. Ever since the first experimental demonstration \cite{PhysRevLett.104.010503}, many proposals developed for constructing CNOT gates
depend on excitation blockade together with a piecewise pulse sequence \cite{PhysRevLett.119.160502}. Such as in the pioneer work \cite{PhysRevLett.104.010503}, authors used seven pulses for the AS-CNOT gate and five pulses for the $H$-C$_Z$-CNOT gate. Subsequent schemes for AS-CNOT gates could work with three laser pulses \cite{PhysRevApplied.9.051001}, even in the weak interaction regime without any rotation errors \cite{PhysRevA.104.012615,PhysRevLett.124.033603}. Other achievements towards $H$-C$_Z$-CNOT gates used an ``one-step implementation" to form a fast $C_Z$ gate \cite{PhysRevApplied.13.024059,PhysRevA.98.032306,Yin:20,Wu:21}, but requiring extra single-qubit Hadamard operations which render the total gate duration elongated \cite{PhysRevLett.123.170503}. More recently, the transition slow-down effect can lead to an accurate AS-CNOT gate with a sub-microsecond duration by two pulses while still working in the strong blockade regime \cite{PhysRevApplied.14.054058}.

In the present work, we analyze a fast and high-fidelity AS-CNOT(briefly CNOT in the following) gate with fewer optical pulses. It works in the van der Waals({\it vdWs}) regime between two isolated Rydberg atoms prepared in a well-defined quantum state. This can essentially avoid the influence of Rydberg blockade.
Compared to the Rydberg gates that work with strong dipole-dipole interactions, the weak {\it vdWs} interaction will not cause the coupling of adjacent two-atom Rydberg levels due to the non-resonant interactions, so as to be less technically demanding. {\it e.g.} for $|62D_{3/2}\rangle$ state in $^{87}$Rb atoms the interaction character of $1/R^6$ cannot be altered when the interatomic distance is $R>5.5$ $\mu$m \cite{PhysRevLett.110.263201}.
 Therefore, accounting for the use of simultaneous excitation with the least number of optical pulses, our protocol not only removes the need for multiple switchings of external control fields reducing the complexity of experimental implementation; but also decreases extra atomic losses induced by the multi-laser noises  \cite{PhysRevLett.121.123603}.

In addition, the scheme has the added advantage of modulating the pulse waveform accurately by {\it optimization algorithm}. We apply a global optimization to
construct amplitude and phase-modulated pulses which implement the two-qubit CNOT or three-qubit Toffoli gates. The pulses are optimized through a genetic algorithm \cite{Mitchell1998An}.
The resulting gate fidelity is essentially constrained by the precision of optimization algorithm and the Rydberg decay error in the weak interaction regime. To make the scheme more accessible, other technical errors arising from the fluctuations of interaction, the laser amplitude noise as well as 
the thermal motion of atoms at finite temperature, are also investigated. In contrast to previous multi-pulse piecewise protocols, our gate benefiting from a simplified pulse design, 
is more straightforward for experimental demonstration and deserves future studies in diverse systems such as superconducting circuits \cite{Devoret1169}, trapped ions \cite{PhysRevLett.91.157901}, and solid-state quantum systems \cite{Zhou:16}.

\section{Theoretical strategy}

\begin{widetext}

\begin{figure}
\includegraphics[width=0.7\textwidth]{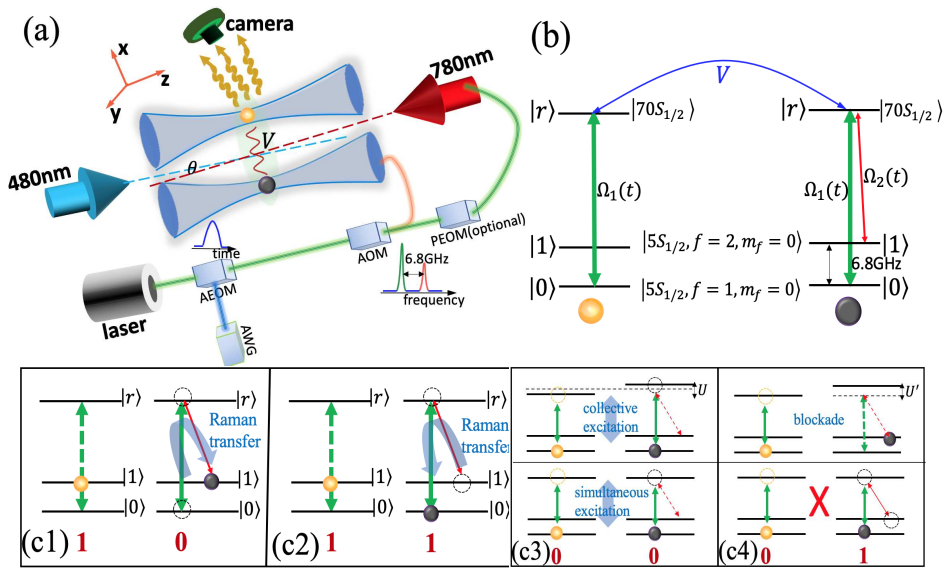} 
\caption{(a) Experimental sketch for a two-qubit CNOT gate. Optical-tweezers-trapped atoms are globally excited by a two-photon excitation with 480nm and 780nm lasers, which propagate in opposite directions with a tiny separation angle $\theta<5^{o}$ for reducing the Doppler-induced frequency shift. Atom measurement can be realized by collecting fluorescence signals on a cooled EMCCD camera placed perpendicularly to the laser propagation. To carry out the gate, several electro-optic modulator(EOM) and acoustic-optical modulator(AOM) devices are needed in experiment. After producing a shaped pulse via an amplitude-EOM(AEOM) one may use an AOM for generating a frequency difference between the zeroth- and first-order diffraction optical fields with respect to the 780nm laser. This light is used for an off-resonant excitation between the ground state $|0\rangle$ or $|1\rangle$ and the intermediate state $|5P_{3/2}\rangle$ (not shown). The other 480nm beam is left on for coupling $|5P_{3/2}\rangle$ to the Rydberg state $|r\rangle=|70S_{1/2}
\rangle$ \cite{2008}. A phase-EOM(PEOM) can selectively provide another phase control between the two pulses. (b) Relevant energy levels including $\{|0\rangle,|1\rangle,|r\rangle\}$ as well as the atom-light couplings. \textcolor{black}{The {\it vdWs} interaction will lead to a pure energy shift $V$ to the two-atom state $|70S_{1/2};70S_{1/2}\rangle$ excluding the influence of other nearby high-lying Rydberg states}.
(c1-c4) Basic physical mechanism provided by a reduced two-pulse protocol, in which the population transformation based on four computational basis states $\{|10\rangle,|11\rangle,|00\rangle,|01\rangle\}$ are shown. Each atom contains $\{|0\rangle$, $|1\rangle$, $|r\rangle\}$ three states.
The modified pulse 1 labeled by the two-photon Rabi frequency $\Omega_1(t)$ (green arrow) globally couples $|0\rangle$ and $|r\rangle$ for both atoms. The local addressing between $|1\rangle$ and $|r\rangle$ of the target atom is provided by pulse 2 with the Rabi frequency $\Omega_2(t)$ (red arrow). 
Solid-arrow stands for a true excitation accompanied by the population transfer, while the dashed-arrow means a null excitation despite the presence of optical pulses.}

\label{modelthy} 
\end{figure}

\end{widetext}

\subsection{Model and Hamiltonian}


To our knowledge, earlier proposals for realizing {\it e.g.} $H$-C$_Z$-CNOT often constructed the entangling gates with a piecewise pulse sequence, accompanied by a local $\frac{\pi}{2}$-rotation which can be implemented via a microwave field \cite{PhysRevA.92.022336}. However, the gate fidelity depends on the phase difference between the microwave field and the optical field, which is usually difficult to be locked in experiments. Besides, the combination of microwave and optical pulses makes the addressing of individual atoms difficult during the gate operation \cite{yan}.
To solve these problems,  we propose a full-optical and straightforward scheme to implement a CNOT gate as illustrated in Fig.\ref{modelthy}(a), which consists of two $^{87}$Rb atoms with a well-separated spacing $r_0$ between the center of two traps (we also study the three-qubit Toffoli gate in Sec. V). The simplified implementation is achieved via a combination of a global coupling $|0\rangle\to|r\rangle$ of both atoms with the effective Rabi frequency $\Omega_1(t)$(green arrow, pulse 1), and the local addressing by a laser inducing the $|0\rangle\to|r\rangle$ transition of the target atom with $\Omega_2(t)$ (red arrow, pulse 2). Atoms are pre-pumped into $|0\rangle$ by a polarized laser propagating along $-\hat{x}$ axis and $|1\rangle$ is a nearby hyperfine ground state. Additional $\pi$ pulses are applied to either or both of the atoms to accomplish the state preparation \cite{PhysRevLett.96.063001}. Two laser pulses differ in frequency by a magnitude of $\omega_{01}/2\pi\approx$ 6.83 GHz which agrees with the hyperfine energy separation between states $|0\rangle$ and $|1\rangle$. In experiment an AOM can be used to spatially split the laser into the zeroth-order (pulse 1) and first-order (pulse 2) fields via diffraction, in which its modulation frequency equals $\omega_{01}$.
A phase modulation with the help of a PEOM is sometimes required for the optimal pulse shapes.


The desired probability matrix for constructing a two-qubit CNOT gate should contain a unitary transformation of the computational basis states $\{|00\rangle,|01\rangle,|10\rangle|11\rangle\}$ as follows:
\begin{equation}
    U_{\text{CNOT}}=\left( 
\begin{array}{cccc}
1 & 0 & 0 & 0 \\ 
0 & 1 & 0 & 0 \\ 
0 & 0 & 0 & 1 \\   
0 & 0 & 1 & 0 \\
\end{array}%
\right)  \label{Ucot} \\ 
\end{equation}
which is enabled by the two-atom Hamiltonian $\hat{\mathcal{H}}=\hat{\mathcal{H}_{c}}\otimes\hat{{I}}+\hat{{I}}\otimes\hat{\mathcal{H}_{t}}
+V\left\vert rr\right\rangle\left\langle rr\right\vert$ with
\begin{eqnarray}
\hat{\mathcal{H}_{c}}&=& \frac{\Omega_{1}(t)}{2}(\left\vert 0\right\rangle \left\langle r\right\vert +H.c.)\nonumber\\
 \hat{\mathcal{H}_{t}} &=& \frac{\Omega_{1}(t)}{2}(\left\vert 0\right\rangle \left\langle r\right\vert +H.c.) +\frac{\Omega_{2}(t)}{2}(\left\vert 1\right\rangle \left\langle r\right\vert +H.c.) \nonumber
\end{eqnarray}
for describing the atom-light interactions of control ($\hat{\mathcal{H}_{c}}$) and target ($\hat{\mathcal{H}_{t}}$) atoms, respectively. $U=V\left\vert rr\right\rangle\left\langle rr\right\vert$ denotes the interatomic Rydberg interaction.

To facilitate the scheme, the Rydberg interaction is assumed to be of {\it vdWs}-feature appearing between two isolated atoms, which excludes the couplings between $ss$ and other higher angular momentum states such as $p$-orbital states \cite{Singer_2005}. More precisely we have summed up all non-resonant dipole-dipole interactions among Zeeman sublevels due to the larger F\"{o}rster defects, leading to a second-order {\it vdWs} shift to the two-atom state $|70S_{1/2};70S_{1/2}\rangle$ \cite{PhysRevA.77.032723}. This differs from the Rydberg phase-shift gates proposed in \cite{PhysRevLett.85.2208,COZZINI2006375}, where the large Rydberg interaction is provided by a static electric field to remove the degeneracy of high-lying Rydberg levels.
We will verify the existence of the {\it vdWs} interaction for appropriate two-atom distances $r_0$ in Appendix A.



\subsection{Physical realization}

Before we proceed to a detailed numerical discussion we consider the theoretical feasibility of the gate implementation. Ideally, the gate operation can be simply understood by two optical pulses $\Omega_1(t)$ and $\Omega_2(t)$, satisfying
\begin{equation}
    |\Omega_1(t)|>|\Omega_2(t)|.
\end{equation}
This assumption is given for simplifying the following discussion but is not crucial for the realization of the gate. The systematic dynamics can be understood by considering the behavior of the four computational basis states:

 (i) If the input state is $|1\beta\rangle$ with $\beta\in\{0,1\}$, {\it i.e.} the control atom is an idler, the two pulses $\Omega_{1}(t)$ and $\Omega_{2}(t)$ transfer the target atom through a two-photon Raman process, obeying
\begin{equation}
   |10\rangle\rightleftarrows|1r\rangle\rightleftarrows|11\rangle   
\end{equation}
[see Fig.\ref{modelthy}(c1-c2)]. Since the control atom is never excited, there is no Rydberg interaction on it leading to null rotation error. The major intrinsic errors stem from the transient residence on $|r\rangle$ of the target atom, {\it i.e.} the Rydberg decay error, as well as the imperfect Raman transfer due to the accuracy of optimization. To minimize the decay error we apply the Raman pulses within a relatively short duration $\sim 1.0$ $\mu$s (Ref.\cite{PhysRevA.101.062309} showed a similar C$_z$-gate duration with optimized adiabatic pulses). Also, the insufficient pulse optimization will bring extra {\it optimization error} here.

(ii) If the control atom is initially in $|0\rangle$ the traditional multi-pulse blockade gate relies on a pre-excitation of the control atom. Once the control atom has been excited, it causes an inevitable energy shift to state $|r\rangle$ of the target atom, preventing its excitation. However, we notice that two pulses are applied simultaneously arising two competitive excitation channels that strongly depend on the interaction strength. With a larger interaction strength, $V>|\Omega_1|,|\Omega_2|$, the $|00\rangle$ state tends to have a collective transition [see (c3), top]
\begin{equation}
    |00\rangle\to\frac{|0r\rangle+|r0\rangle}{\sqrt{2}}\to|00\rangle.
    \label{way1}
\end{equation}
This channel only involves single-atom Rydberg excitation so as to be insensitive to the fluctuation of interactions.
Oppositely, when the interaction strength is small, $V\leq|\Omega_1|,|\Omega_2|$, it facilitates a simultaneous two-atom excitation [see (c3), bottom] 
\begin{equation}
    |00\rangle\to|rr\rangle\to|00\rangle.
    \label{way2}
\end{equation}
This is a {\it which-way} problem. The realistic excitation channel is interaction-dependent, so no matter big or small the interaction strength is, our scheme can apply targeted approaches.
However, for the target atom initially in $|1\rangle$ [see (c4)] we are interested in the regime $|\Omega_1|>|\Omega_2|$, in which the weak coupling $\Omega_2(t)$ is unable to overcome the energy shift induced by the {\it vdWs} interaction, since the control atom is pre-excited by $\Omega_1(t)$. As a result, the target atom is left on $|1\rangle$ without excitation, giving rise to an ideal transfer of
\begin{equation}
   |01\rangle\to|r1\rangle\to|01\rangle.
     \label{way3}
\end{equation}
Such a process may suffer from an imperfection. When the shifted energy is small leading to an excitation of the target atom to $|r\rangle$ by $\Omega_2(t)$, the subsequent coupling between $|r\rangle$ and $|0\rangle$ by $\Omega_1(t)$ would destroy the targeted state transformation by impelling the target atom onto an undesired state $|0\rangle$, as shown in the lower panel of (c4).

It should be stressed that we work on the implementation of two- and three-qubit CNOT gates by simply placing them in the {\it vdWs}-interaction regime. It is challenging to construct a global optimization to the amplitude and phase of the two pulses simultaneously. Especially a weak two-atom interaction will make the condition $|\Omega_1(t)|>|\Omega_{2}(t)|$ unsatisfied. In that case, the complicated atomic dynamics can not be easily understood by the map of Fig.\ref{modelthy}(c). However, with a robust optimization of the pulse parameters, we can always guarantee the state transformation of the system with a high fidelity.

\section{Gate performance}

\subsection{Two-pulse scheme}

To achieve desired optical pulses that work on all input states we use the {\it genetic algorithm} to optimize an amplitude and phase profile for the driving laser fields. 
The principle of optimization algorithm is given in Appendix B. In the procedure,
 we treat the gate fidelity  $\mathcal{F}_2$ (see Eq.(\ref{fidelity}) where the subscript means the number of qubits) as a single objective function, by which the gate error denoted by $1-\mathcal{F}_2$ attains its global minimum under multi-time optimization. We begin with a guess for the waveforms of two optical pulses, forming of an approximate Fourier-series expansion by taking account of the symmetry of state transformation \cite{Kido2015}:
\begin{eqnarray}
\Omega_1(t) = \Omega_{10}+\Omega_{11}\cos(2\pi t/T_g)+\Omega_{12}\sin(\pi t/T_g)  \nonumber \\
\Omega_2(t) = \Omega_{20}+\Omega_{21}\cos(2\pi t/T_g)+\Omega_{22}\sin(\pi t/T_g) \nonumber
\end{eqnarray}
where the optimized pulse has a duration of $T_g=1.0$ $\mu$s. 
In order to meet the single constraint $|\delta\mathcal{F}_2|< 10^{-4}$, where $|\delta\mathcal{F}_2|$ is the fluctuation of fidelity, we optimize the pulse parameters $(\Omega_{10},\Omega_{11},\Omega_{12})$ and $(\Omega_{20},\Omega_{21},\Omega_{22})$ at the same time. Experimental setup utilizing an AEOM to shape optical pulses, has been carried out on sub-microsecond timescales in a solid-state quantum system \cite{Zhou:16}. An extra phase modulation by PEOM is performed on one of them accompanied by a path locking technique, which finally enables an arbitrary phase difference between the two pulses.

\begin{figure}
\includegraphics[width=0.5\textwidth]{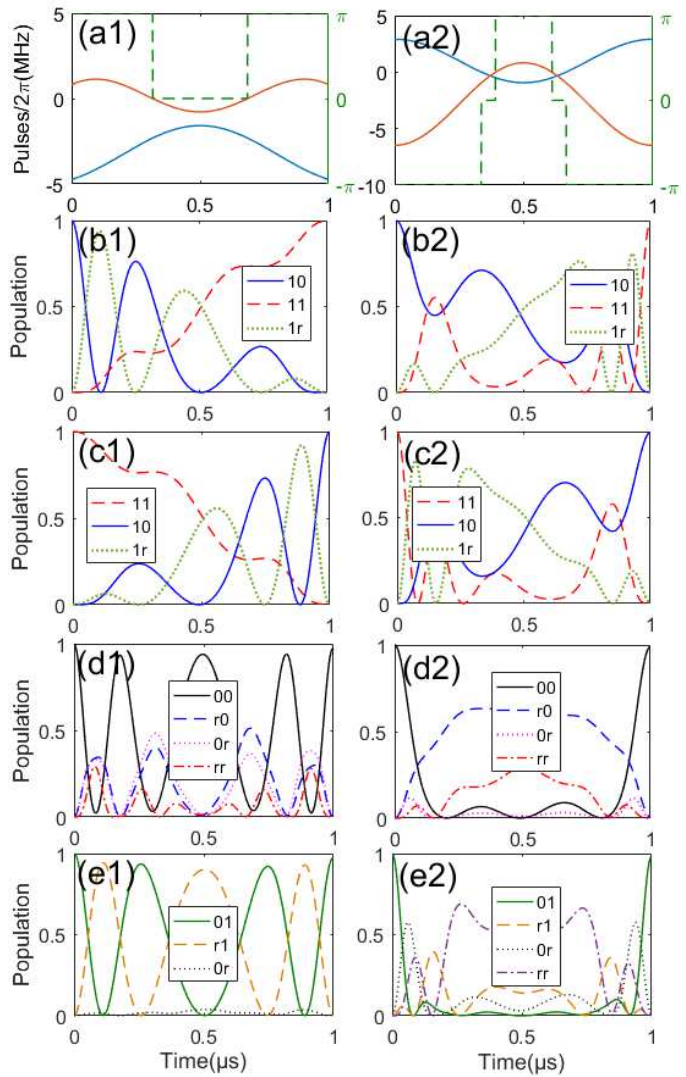} 
\caption{From top to bottom, (a1-a2) two optimal ground-Rydberg pulses $\Omega_1(t)$(blue), $\Omega_{2}(t)$(red). The green-dashed curve shows the phase difference $\delta\phi=\phi_1-\phi_2$ between two pulses and a precise phase control is needed; (b1-e2) the real population dynamics under the inputs of four computational basis states $\{|10\rangle$, $|11\rangle$, $|00\rangle$, $|01\rangle\}$. Different central interactions $V_0/2\pi$=7.0 MHz($r_0=7.10$ $\mu$m) for (a1-e1) and $V_0/2\pi$=1.0 MHz($r_0=9.76$ $\mu$m) for (a2-e2), are used excluding the fluctuation of interactions.}
\label{optimalpulses} 
\end{figure}

 \begin{table*}
\caption{\label{tab:table1}%
Optimized pulse amplitudes and the gate fidelity $\mathcal{F}_2$ excluding the fluctuations to the interatomic interaction.
All coefficients $(\Omega_{10},\Omega_{11},\Omega_{12})$ and $(\Omega_{20},\Omega_{21},\Omega_{22})$ are obtained under multi-optimization until the required precision $|\delta \mathcal{F}_2|<10^{-4}$ can be reached. The {\it vdWs} dispersion coefficient is $C_6/2\pi = 863$ GHz$\cdot \mu m^6$ for state $|70S_{1/2};70S_{1/2}\rangle$ and $V_0=C_6/r_0^6$ represents the non-fluctuated {\it vdWs} interaction between two distant atoms.}
\begin{ruledtabular}
\begin{tabular}{cccccc}
$r_{0}$ ($\mu m$)&$V_{0}/2\pi$ (MHz)&$\gamma$ (kHz)&($\Omega_{10}$,$\Omega_{11}$,$\Omega_{12})$/2$\pi$ (MHz)&($\Omega_{20}$,$\Omega_{21}$,$\Omega_{22})$/2$\pi$ (MHz)&$\mathcal{F}_2$\\
\hline
7.10& 7.0&3.0& (-3.9621, -0.7858, 1.5915)&(1.0942, -1.9068, -2.2182)&0.9921\\
9.76& 1.0&3.0& (-0.9549, -1.9544, -0.0631)&(-2.8310, -3.6488, 0.0074)&0.9951\\
\end{tabular}
\end{ruledtabular}
\end{table*}

To evaluate the fidelity, we first numerically
solve the population dynamics by integrating the two-atom master equation \cite{PhysRevA.95.012708}
\begin{equation}
   \frac{d\hat{\rho}}{dt}=-i[\hat{\mathcal{H}},\hat{\rho}]+\hat{\mathcal{L}}[\rho]
   \label{rhoeq}
\end{equation}
 with $\hat{\mathcal{H}}$ the two-atom Hamiltonian. The radiative decay term of the Lindblad operator is given by
\begin{equation}
    \hat{\mathcal{L}}[\rho]=\sum_{j=1}^{4}[\hat{L}_j\hat{\rho}\hat{L}_j^{\dagger}-\frac{1}{2}(\hat{L}_j^{\dagger}\hat{L}_j\hat{\rho}+\hat{\rho}\hat{L}_j^{\dagger}\hat{L}_j)]
\end{equation}
where
\begin{eqnarray}
\hat{L}_1=\sqrt{\gamma}[|10\rangle\langle r0|+|11\rangle\langle r1|+|1r\rangle\langle rr|] \nonumber \\
\hat{L}_2=\sqrt{\gamma}[|00\rangle\langle r0|+|01\rangle\langle r1|+|0r\rangle\langle rr|] \nonumber \\
\hat{L}_3=\sqrt{\gamma}[|01\rangle\langle 0r|+|11\rangle\langle 1r|+|r1\rangle\langle rr|] \nonumber \\
\hat{L}_4=\sqrt{\gamma}[|00\rangle\langle 0r|+|10\rangle\langle 1r|+|r0\rangle\langle rr|] \nonumber
\end{eqnarray}
separately demonstrate the spontaneous decay ($\gamma$ is the Rydberg decay rate) of $|r\rangle\to|1\rangle$ and $|r\rangle\to|0\rangle$ of control($\hat{L}_{1,2}$) and target($\hat{L}_{3,4}$) atoms. 
Relying on the mature technology for cooling and manipulating individual $^{87}$Rb atoms, we adopt $\gamma = 3.0$ kHz in a cryogenic environment for states $|r\rangle=|70S_{1/2}\rangle$, $|0\rangle=|5S_{1/2},F=1,m_F=0\rangle$, $|1\rangle=|5S_{1/2},F=2,m_F=0\rangle$ in which the intermediate state $|5P_{3/2}\rangle$ is ignored for a two-photon process (note that this is a little different from the experimental setup in Ref.\cite{PhysRevLett.123.170503} where the intermediate state is $|6P_{3/2}\rangle$). The observable function for the two-qubit gate fidelity is defined as
\begin{equation}
\mathcal{F}_2=\frac{1}{4}Tr\left[\sqrt{\sqrt{U_\text{CNOT}}\rho_{det}(t=T_g)\sqrt{U_\text{CNOT}}}\right]
\label{fidelity}
\end{equation}
by taking account of the average effect of four computational basis states. $\rho_{det}$ represents the detected density matrix at $t=T_g$ that evolves according to Eq.(\ref{rhoeq}). We adopt numerical optimization to construct amplitude and phase modulated pulses by which
the system evolves satisfactorily, realizing a high-fidelity CNOT gate.

To demonstrate the detailed dynamics of the system with respect to Eq.(\ref{rhoeq}) we first present the numerical results in Fig.\ref{optimalpulses} without fluctuated interactions. For completeness, we give the corresponding numerical values of relevant parameters in Table I.
Under the modulated pulses as observed in Fig.\ref{optimalpulses}(a1-a2), the dynamics of $|10\rangle$($|11\rangle$) are given by the coupling of the target atom on the $|0\rangle(|1\rangle)\rightleftarrows|r\rangle\rightleftarrows|1\rangle(|0\rangle)$ transition, forming a two-photon Raman transfer [Fig.\ref{optimalpulses}(b1-c2)].
That means except the Rydberg decay error, the major intrinsic error is related to the transferring of the target atom, which comes from the imperfect pulse optimization. For a chosen value of interaction $V_0$, the dynamics behave quite differently. When $V_0$ is relatively large [$V_0/2\pi=7.0$ MHz, (b1)-(c1)] it arises $|\Omega_1(t)|>|\Omega_2(t)|$ under the optimization. 
For the dynamics associated with $|10\rangle$ or $|11\rangle$, the population transfers as expected, especially the Rabi oscillation between $|10\rangle$ and $|1r\rangle$ is stronger than that between $|1r\rangle$ and $|11\rangle$. This contrast is attributed to $|\Omega_1(t)|>|\Omega_2(t)|$. 
A small $V_0$ [$V_0/2\pi=1.0$ MHz, (b2)-(c2)] leads to the comparable amplitudes of two optimized pulses. The dynamics is a bit complicated in this case, but a satisfactory population transfer is still achievable accompanied by irregular oscillations.

We proceed to study a more complex case if the input is $|00\rangle$ or $|01\rangle$ which is sensitive to the interaction. Accounting for the {\it which-way} mechanism two atoms with different $V_0$ will have different excitation channels such as Eq.(\ref{way1}) and Eq.(\ref{way2}) even though the initial state is same. For example, see (d1), if $V_0/2\pi=7.0$ MHz leading to $|\Omega_1|,|\Omega_2|<V_0$, due to the {\it vdWs} blockade mechanism \cite{PhysRevLett.99.163601,PhysRevLett.112.183002}, the state $|00\rangle$ evolves to the middle singly-excited states $|r0\rangle$ and $|0r\rangle$, 
and hence the scheme is not sensitive to the fluctuation of interactions. In contrast, see (d2), when $V_0/2\pi=1.0$ MHz and then $|\Omega_1|,|\Omega_2|\geq V_0$, we observe that the condition $|\Omega_1|>|\Omega_2|$ is not satisfied. The mixture of various singly-excited and doubly-excited Rydberg states causes unpredictable population dynamics. The latter process involving dominant double Rydberg excitation would be sensitive to the variation of atomic spacing. For the dynamics associated with state $|01\rangle$[(e1) and (e2)] similar behaviors are observed, but the only case (e1) shows an ideal dynamical evolution for the control atom between $|01
\rangle$ and $|r1\rangle$ [see Eq.(\ref{way3})]. As a consequence, the optimal pulse amplitudes are mainly determined by the interaction energy $V_0$ which arises a changeable dynamics, even if the input is same.

In general, a large interaction energy with $V_0>|\Omega_1|>|\Omega_2|$ can trigger a finite excitation blockade in the {\it vdWs} regime, allowing one of the atoms to be excited at any time (Fig.2(e) of Ref.\cite{PhysRevLett.110.263201} showed a collective single-atom excitation with an enhanced Rabi frequency for $|62D_{3/2}\rangle$ of rubidium atoms when $r_0$ is about 4.0 $\mu$m). While a weaker interaction makes the atoms redistributed in various energy levels. The simulated gate fidelity including intrinsic decay and optimization error, 
is $\mathcal{F}_2=0.9921$ for $V_0/2\pi=7.0$ MHz and $\mathcal{F}_2=0.9951$ for $V_0/2\pi=1.0$ MHz.

\subsection{Fluctuation of interactions}

\begin{figure}
\includegraphics[width=0.48\textwidth]{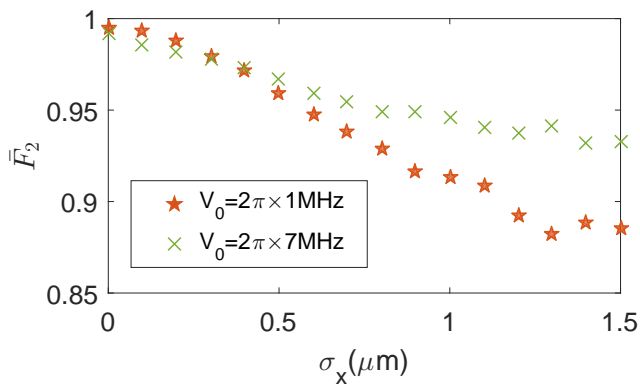} 
\caption{Average gate fidelity $\bar{\mathcal{F}}_2$ as a function of the standard deviation $\sigma_x$ for different central interactions $V_0/2\pi=1.0$ MHz(stars), 7.0 MHz(crosses). During each measurement the real interatomic distance $r$ is random satisfying a three-dimensional Gaussian distribution with mean values $x_0=(C_6/V_0)^{1/6}$, $y_0=z_0=0$ and standard derivations $\sigma_{x,y,z}$.
Here $\sigma_y=\sigma_z=0.27$ $\mu$m and $\sigma_x$ can be widely adjusted.
Each point denotes an average of 500 measurements. Relevant optimal parameters basing on no fluctuated interactions are given in Table I.}
\label{sigmaxvar} 
\end{figure}

In fact, the real interatomic distance $r$ cannot be strictly fixed during the gate operation, leading to the fluctuation of interactions. In the numerical simulation, we model the distance $r$ by a three-dimensional Gaussian distribution with mean values $x_0=(C_6/V_0)^{1/6}$, $y_0=z_0=0$.
The distance errors(standard deviation) are estimated to be $\sigma_{y}=\sigma_{z}=0.27$ $\mu$m and $\sigma_x\in[0,1.5]$ $\mu$m is a tunable value, which agree with the experimental results reported in Ref. \cite{PhysRevLett.123.230501}.
 During each measurement we extract a random position $r_i$ according to the Gaussian distribution, giving rise to a fluctuated interaction $V_i$. By applying $V_i$ instead of $V_0$, we re-calculate the gate fidelity $\mathcal{F}_{2,i}$ for the $i$th measurement. The final average fidelity is provided by
\begin{equation}
  \bar{\mathcal{F}}_2=\sum_{i=1}^{\mathcal{N}}{\mathcal{F}_{2,i}}/\mathcal{N}.
\end{equation}
Figure \ref{sigmaxvar} presents the $\bar{\mathcal{F}}_2$ as a function of $\sigma_x$ under a sufficient number of measurements, $\mathcal{N}=500$. Since all pulse parameters have been optimally obtained under a non-fluctuated interaction $V_0$, they are unable to be exactly appropriate for the real population evolution with a fluctuated interaction $V_i$. Hence the reduction of $\bar{\mathcal{F}}_2$ with respect to $\sigma_x$ is clearly observed no matter what the $V_0$ is. Besides owing to the suppression of double Rydberg excitation by a finite {\it vdWs} energy shift, in the case of a larger $V_0$,
the gate fidelity $\bar{\mathcal{F}}_2$ shows a good robustness to the increase of $\sigma_x$. It is apparent that the average fidelity can maintain a higher value, $\bar{\mathcal{F}}_2\approx0.9327$, even at $\sigma_x=1.5\mu$m for $V_0/2\pi=7.0$ MHz. Yet this value will reduce to 0.8857 for $V_0/2\pi=1.0$ MHz.

\subsection{Improved one-pulse scheme}

\begin{figure}
\includegraphics[width=0.5\textwidth]{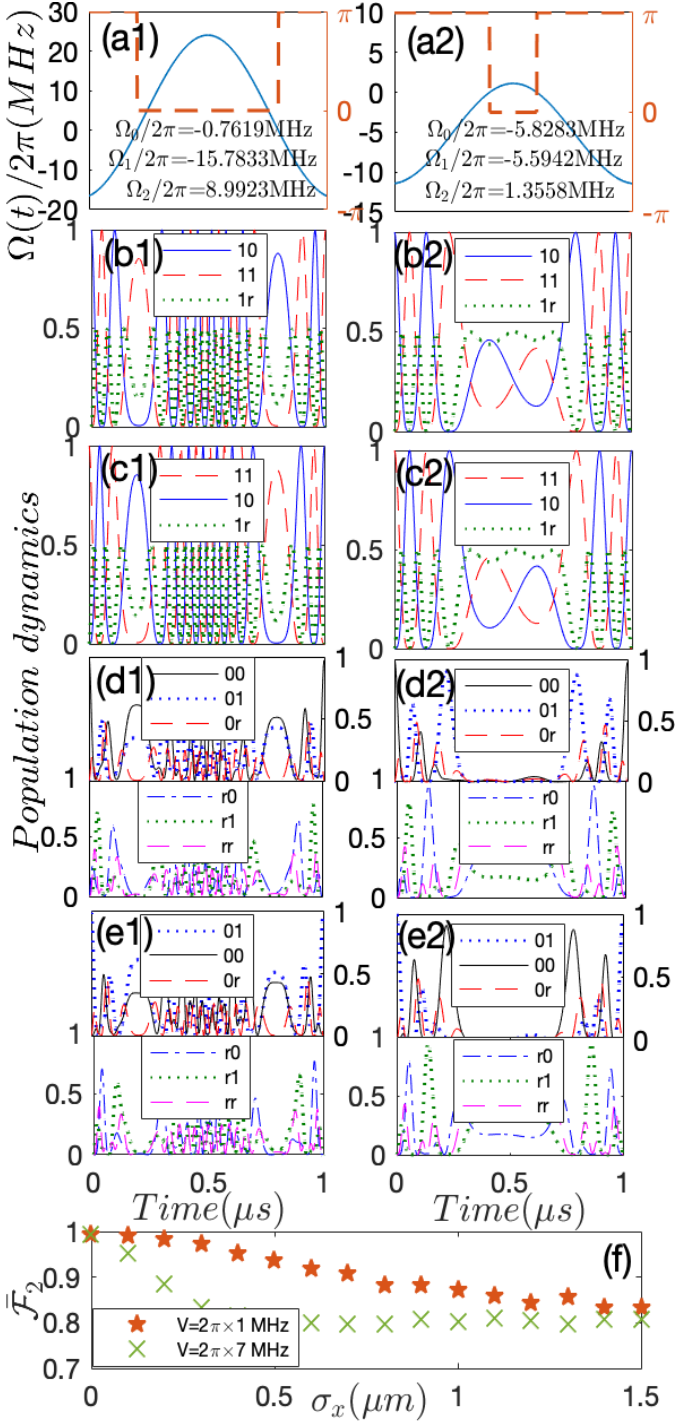} 
\caption{Realization of an one-pulse two-qubit CNOT gate. From top to bottom, (a1) the optimized pulse shape(blue-solid) and phase variation(red-dashed) under $V_0/2\pi=7.0$ MHz. Coefficients $(\Omega_0,\Omega_1,\Omega_2)$ are explicitly shown in the picture. (b1-e1) Numerically-simulated population dynamics of four computational basis states $\{|10\rangle,|11\rangle,|00\rangle,|01\rangle\}$ excluding the effect of fluctuated interactions.
Similarly, (a2-d2) show the results for $V_0/2\pi=1.0$ MHz. (f) Average fidelity $\bar{\mathcal{F}}_2$ with respect to the variation of the standard deviation $\sigma_x$ for $V_0/2\pi=$1.0 MHz(stars), 7.0 MHz(crosses). Each point denotes an average of 500 measurements.}
\label{onep} 
\end{figure}

To obtain a minimal model for the targeted gate the two optical pulses can even have same amplitudes generated by one AEOM (see Fig.\ref{modelthy}(a) and Sec.VI for more details). This so-called {\it one-pulse scheme} only requires one optical pulse taking a similar form of
\begin{equation}
    \Omega(t) = \Omega_{0}+\Omega_{1}\cos(2\pi t/T_g)+\Omega_{2}\sin(\pi t/T_g).
    \label{Ome}
\end{equation}
We seek appropriate values $(\Omega_0,\Omega_1,\Omega_2)$ for achieving a satisfactory gate performance.
During the gate implementation, the global coupling over both qubits between $|0\rangle$ and $|r\rangle$, as well as the local transition from $|1\rangle$ to $|r\rangle$ for the target qubit, are played by a same-shaped pulse with the Rabi frequency $\Omega(t)$. The laser is frequency modulated by an AOM to produce the zeroth- and first-order diffractions which achieves a spatial separation for driving two separated atomic qubits. The {\it one-pulse scheme} can strongly prohibit the unwanted phase control between the two pulses, promising for the true realization of a minimal two-qubit CNOT gate.

Simulation in Figure \ref{onep} shows the population dynamics of four states $\{|10\rangle,|11\rangle,|00\rangle,|01\rangle\}$ under an optimized pulse.
After an optimization with $T_g=1.0$ $\mu$s we reach an amplitude and phase modulated pulse as displayed in (a1-a2). We note that a peak Rabi frequency of similar magnitude was used for implementing the Rydberg adiabatic gates \cite{PhysRevA.89.030301,PhysRevX.10.021054}.
The dynamics of $|10\rangle$ (and $|11\rangle$) is affected by the $|0\rangle\to|r\rangle\to|1\rangle$(and $|1\rangle\to|r\rangle\to|0\rangle$) transition of the target atom, forming a three-level coupled system. The resulting coherent population conversion between two ground states $|10\rangle$ and $|11\rangle$ shows strong oscillations through a two-photon process mediated by $|1r\rangle$. Also, due to the symmetric waveform of $\Omega(t)$ the dynamics of $|10\rangle$ and $|11\rangle$ share a same behavior as time going [see (b1-c1) and (b2-c2)].
The population evolution of state $|00\rangle$ ($|01\rangle$) strongly depends on the strength of the interaction because of the {\it which-way} mechanism, which leads to a wide distribution of populations on various middle Rydberg states including $|0r\rangle$, $|r0\rangle$, $|rr\rangle$, and $|r1\rangle$. Finally, while excluding the fluctuation of interactions, we show that
the one-pulse two-qubit CNOT gate realized has a fidelity of $\mathcal{F}_2\sim0.9923$ for $V_0/2\pi=7.0$ MHz and $\mathcal{F}_2\sim0.9935$ for $V_0/2\pi=1.0$ MHz in a cryogenic environment.

To obtain more practical numbers we also add extra fluctuations to the interaction $V_0$ characterized by position error $\sigma_{x,y,z}$ in the three spatial dimensions. As shown in Fig.\ref{onep}(f), $\sigma_{x}$ is changed while $\sigma_y=\sigma_z=0.27$ $\mu$m. The simulation of the average fidelity $\bar{\mathcal{F}}_2$ shows a similar tendency as in the two-pulse case where $\bar{\mathcal{F}}_2$ decreases as $\sigma_x$ grows. Here the scheme with $V_0/2\pi=7.0$ MHz is more sensitive to the fluctuations of interaction, because the fast-oscillating excitation of $|rr\rangle$ substantially reduces the gate fidelity. So taking into account the position error with $\sigma_x=1.5$ $\mu$m, $\sigma_{y(z)}=0.27$ $\mu$m,
the resulting average fidelity $\bar{\mathcal{F}}_2$ is decreased to $\sim0.8043$ for $V_0=7.0$ MHz and $\sim0.8322$ for $V_0=1.0$ MHz after 500 measurements.

We demonstrate an unprecedented realization of a two-qubit CNOT gate with only one optical pulse, which is enabled by a robustly-optimized pulse waveform that works for all input states. Different from in a two-pulse scheme, it is not required here to face the technical difficulties coming from a careful alignment of the two pulses on the time axis and a precise control of their phase difference. Therefore, our scheme can offer a promising route to implement a fast and convenient two-qubit CNOT gate.
On the other hand, a pure {\it vdWs} interaction is assumed between two well-separated qubits, instead of a strong dipole-dipole interaction, which immunizes the gate to any leakage error from nearby Rydberg states. A detailed proof for a pure {\it vdWs} interaction when $r_0=7.10$ $\mu$m or $9.76$ $\mu$m is given in Appendix A.


\section{Error estimates}

\subsection{Intrinsic errors}

{\it Rydberg decay error.} Except for the position error characterized by standard deviations $\sigma_{x,y,z}$, the simulated gate fidelity is mostly limited by the Rydberg decay error which increases with the rate of decays. Yet in our scheme for any state dynamics, due to the smooth waveform of the pulse amplitude modulation, the total duration for the Rydberg population of four input states can be reduced. That fact differs from a piecewise C$_z$ gate sequence adopted in a Rydberg-blockade-based CNOT gate, where the control atom persists on the Rydberg state during the pulse duration applied to the target atom, arising a large decay loss. {\it e.g.} in Ref. \cite{PhysRevA.85.042310}, an estimation of decay error shows $\mathcal{E}_{sp}=2.6\times 10^{-3}$ for trapped $97d_{5/2}$($\tau\sim300$ $\mu$s)  rubidium atoms at a temperature of 175 $\mu$K. And this error increases to $7.5\times 10^{-3}$ for $66s$($\tau\sim130$ $\mu$s) cesium atoms in a room temperature \cite{PhysRevLett.123.230501}.

\begin{figure}
\includegraphics[width=0.5\textwidth]{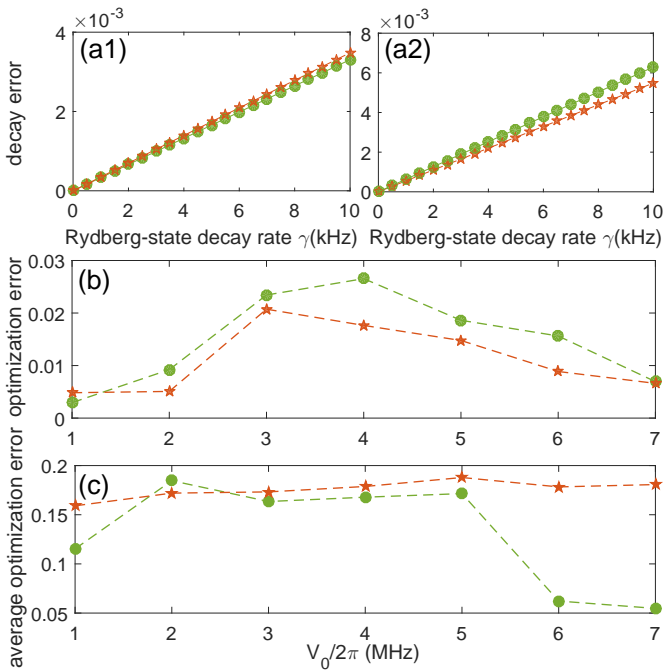} 
\caption{Estimated decay error $\mathcal{E}_{sp}$ {\it vs} the Rydberg decay rate $\gamma$ under (a1) $V_0/2\pi=7.0$ MHz and (a2) $V_0/2\pi=1.0$ MHz. (b) Calculated optimization imperfection $\mathcal{E}_{opt}$ as a function of the {\it vdWs} interaction strength $V_0$ where $\gamma=0$. For each $V_0$ all pulse coefficients have to be re-optimized, see more details given in Table III. (c) Average optimization error $\bar{\mathcal{E}}_{opt}$ under 500 measurements taking account of the fluctuated interactions. Here $\sigma_x=1.5$ $\mu$m, $\sigma_y=\sigma_z=0.27$ $\mu$m.
The results from the one(two)-pulse scheme are marked by red stars(green dots), respectively.}
\label{decayerr} 
\end{figure}



 In our protocol, to estimate the influence of the Rydberg-state decay we numerically compute the gate error with respect to the variation of the decay rate ($\gamma\sim1/\tau$) where $\tau$ is the Rydberg lifetime.
In Fig.\ref{decayerr}(a1-a2), we show the results by shifting the curve starting from zero in order to exclude the influences from other imperfections. For each value $\gamma$ we use same pulse parameters as described in Table I and Fig. \ref{onep}(a1-a2). The fluctuation of the qubit positions  is ignored for showing a pure relationship.
Compared to the Rydberg CNOT gates that work in a strong blockade regime the 
 decay error of our gate has a similar relationship which shows $\mathcal{E}_{sp}\propto\gamma$ (Eq.1 of Ref.\cite{PhysRevA.82.030306} explicitly showed $\mathcal{E}_{sp}\propto1/\tau$ in a controlled-phase gate where $\tau$ is the radiative lifetime of the Rydberg level),
 yet benefiting from a lower value. The reason is that the smooth modulation of the pulse amplitudes keeps small Rydberg population in $|rr\rangle$, $|\beta r\rangle$, $|r\beta\rangle$ with $\beta\in\{0,1 \}$. Based on relevant parameters in \cite{PhysRevA.79.052504}, a numerical estimation arises $\mathcal{E}_{sp}\approx 3.5\times10^{-3}$ ($\mathcal{E}_{sp}\approx 6.0\times10^{-3}$) above the room temperature for $V_0/2\pi=7.0$ MHz ($V_0/2\pi=1.0$ MHz) where $\tau \sim 100$ $\mu$s. When the atom temperature is set to $50$ $\mu$K with $\tau \sim 400$ $\mu$s this error has a clear decrease leading to $\mathcal{E}_{sp}\approx 8\times10^{-4}$ ($\mathcal{E}_{sp}\approx 1.6\times10^{-3}$), accordingly. By comparing these values we conclude that, the spontaneous emission from the Rydberg states is one of the major limitations on the gate fidelity in the weak interaction regime, however it has been well controlled to a compatible level with previous gate protocols.


 {\it Resilience to the change of {\it vdWs} interaction.} Furthermore, we need to explore the influence of different interatomic interactions on the gate performance, which mainly depends on the optimization. That differs from a Rydberg blockade gate where a residual blockade error can be deeply reduced by a huge dipole-dipole interaction between high-lying Rydberg states (The second term in Eq.2 of Ref. \cite{Saffman_2011} showed the intrinsic error of a Rydberg blockade gate is proportional to $ 1/V_0^2$). 
 In other words the gate should work in the strong dipole-dipole interaction regime with a small qubit spacing where its interaction strength is large enough as compared with the peak Rabi frequencies of all driving pulses \cite{PhysRevA.72.022347}. 
 \textcolor{black}{However, our gate works in the weak {\it vdWs}-interaction regime with a limited blockade strength. Owing to the exclusion of nearby two-atom Rydberg states we could study the influences of different interaction strengths based on a pure {\it vdWs} environment.}

  In our scheme, for the dynamics associated with $|10\rangle$ and $|11\rangle$ the double excited Rydberg state $|rr\rangle$ does not receive any population because the control atom in $|1\rangle$ is not coupled by the laser pulses. Yet, the population dynamics with states $|00\rangle$ and $|01\rangle$ are indeed impacted by using different interaction strengths. Luckily, thanks to the {\it which-way} mechanism there always exists a possible route for the evolution of population dynamics no matter what the interaction is. Especially in the case of a weaker interaction all populations are distributed in a number of Rydberg levels [e.g. see Fig.2(d2),(e2)], mixing with multi-transferred routes.
That is the reason why our protocol can show a resilient insensitivity to the change of interaction strengths. The smooth modulation of the pulse amplitudes via optimization helps to suppress the rotation error and keep a minimal population in $|rr\rangle$. Except for the decay error, another intrinsic limitation on our gate fidelity comes from the optimization imperfection due to the limited precision of algorithm, as estimated in Fig.\ref{decayerr}(b). This optimization error $\mathcal{E}_{opt}$ has shown its insensitivity to the change of $V_0$ for both one- and two-pulse schemes.

During a single measurement without interaction fluctuations, the optimization error $\mathcal{E}_{opt}$ can sustain around $\sim 0.01$ for any $V_0$ values. We note that the optimization is more efficient at $V_0/2\pi=1.0$ MHz(smaller) and $7.0$ MHz(larger) because of the {\it which-way} mechanism.
Furthermore, by carrying out sufficient measurements with position fluctuations: $\sigma_x=1.5\mu$m, $\sigma_{y(z)}=0.27\mu$m, we can reach an average optimization error $\bar{\mathcal{E}}_{opt}$ as displayed in Fig.\ref{decayerr}(c). We observe that, in the one-pulse case(red stars), $\bar{\mathcal{E}}_{opt}$ remains to be around 0.15 for any $V_0$ because the population on state $|rr\rangle$ shows little change [see Fig.4(d1-e2)]. 
 While turning to the case of two pulses(green dots), the excitation process in a larger-interaction regime with $V_0/2\pi> 5.0$ MHz acquires a big suppression to the population of $|rr\rangle$, so as to be insensitive to the fluctuation of qubit separation. Thus the $\bar{\mathcal{E}}_{opt}$ can reach as low as {$\sim0.058$}(an average value when $V_0/2\pi=(6.0,7.0)$ MHz), which is much smaller than the average value {$\sim0.179$} given by the one-pulse scheme in the same regime. These results conclude that, provided the spontaneous emission from the Rydberg level has been suppressed via a higher-lying Rydberg level or a cryogenic environment, the intrinsic obstacle for a high gate fidelity is only constrained by the precision of optimization.


\subsection{Technical errors}

\begin{figure}
\includegraphics[width=0.49\textwidth]{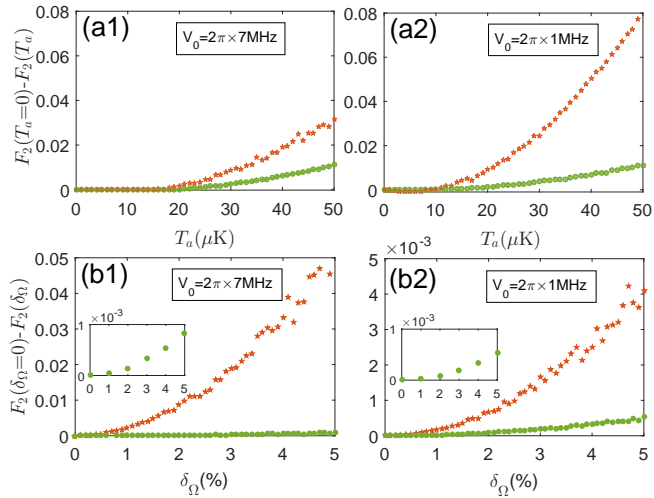} 
\caption{Imperfections of the gate fidelity caused by (a1-a2) the motional dephasing effect under different temperatures $T_a\in[0,50]$ $\mu$K and (b1-b2) the fluctuation of laser amplitudes where $\delta_{\Omega}=\delta\Omega/0.5\Omega_{1(2)}^{pp}$ ($\delta_{\Omega}=\delta\Omega/0.5\Omega^{pp}$ for the one-pulse scheme) stands for the ratio between the maximal deviation $\delta\Omega$ and half of the peak-peak value of optimized pulse amplitudes. Each point denotes an average of 500 samplings. Results from the two-pulse and one-pulse schemes are marked by green dots and red stars respectively without fluctuated interactions. Insets of (b1-b2) amplify the data solved from the two-pulse case. }
\label{temperature} 
\end{figure}

In what follows, we analyze the gate performance due to some technical imperfections, such as the thermal motion of atoms and the fluctuation from laser amplitudes. During the gate operation, the atomic thermal motion at a finite temperature will induce an inevitable Doppler-dephasing to the excitation of the Rydberg state, which can be estimated as a phase change of the two-photon Rabi frequencies in the two-pulse case, obeying 
\begin{equation}
\Omega_{1}\rightarrow\Omega_{1}e^{i\Delta_{1} t},  
\Omega_{2}\rightarrow\Omega_{2}e^{i\Delta_{2} t} 
\end{equation}
Note that this replacement turns to be $\Omega\rightarrow\Omega e^{i\Delta t}$ for the one-pulse scheme.
The detunings $\Delta_{1(2)}$ and $\Delta$ in the phase factor of the Rabi frequencies can be approximately characterized by a Gaussian distribution around its desired value $\bar{\Delta}=0$ with the standard deviation $\sigma_{\Delta}$. Typically, $\sigma_{\Delta}=\boldsymbol{k}\boldsymbol{v}$ where $\boldsymbol{k}=\sum_j \boldsymbol{k}_j$ is the overall wave vector and $\boldsymbol{v}$ is the root-mean-square velocity. Although a Doppler-free three-photon transition was proposed by satisfying $\boldsymbol{k}=0$ with a specific starlike planar geometry of three optical fields \cite{PhysRevA.84.053409}, there is still no way to entirely eliminate the Doppler effect in a two-photon transition which is mostly used. Ref. \cite{PhysRevApplied.13.024008} presents a detailed discussion for how to suppress the Doppler-dephasing induced by atomic thermal motion via two counterpropagating sets of fields. Presently, to minimize this effect, we also use counterpropagating beams at 780nm and 480nm for the transition $|5S_{1/2}\rangle\to |5P_{3/2}\rangle\to|70S_{1/2}\rangle$, leading to $\sigma_{\Delta}=(\boldsymbol{k}_{480}-\boldsymbol{k}_{780})|\boldsymbol{v}|
=k v_{rms}$. Here $k=5\times10^6$ $m^{-1}$ and $v_{rms}=\sqrt{k_BT_a/M}$ where $k_B$ is the Boltzmann constant and $M$ is the atomic mass. 
The atomic temperature $T_a$ is assumed to change within the range of $T_a\in[0,50]$ $\mu$K for a cryogenic environment. We note that similar magnitudes of an atomic temperature have been reached experimentally \cite{PhysRevA.99.043404,2018,2021}.

The imperfection of the overall gate fidelity represented by $\mathcal{F}_{2}(T_a=0)-\mathcal{F}_2(T_a)$ as a function of $T_a$ is shown in Fig. \ref{temperature}(a1-a2). For a given temperature $T_a$ we randomly adopt a detuning from the $T_a$-dependent Gaussian distribution to simulate the influence of phase error on the population evolution. Each point denotes the result obtained by averaging over 500 samplings. The numerical results reveal that the two-pulse scheme has stronger robustness to the thermal-motion induced dephasing. Even at $T_a=50$ $\mu$K corresponding to $\sigma_{\Delta}= 0.3498$ MHz,
the gate imperfection can be suppressed to $\sim 0.0111$ no matter what the interatomic interaction is. On the contrary, the population dynamics of the one-pulse scheme are more sensitive to the extra phase fluctuations of the Rabi frequency, resulting in a larger motion-induced dephasing error.
For example the imperfection of gate fidelity increases to be \textcolor{black}{$\sim0.0318(0.0813)$ for $V_0/2\pi=7.0(1.0)$ MHz at $T_a=50$ $\mu$K}.
In experiment, to suppress the motional dephasing between the ground state and the Rydberg state, one efficient way is to cool atoms via Raman sideband cooling which leads to a temperature about 10 $\mu$K or below \cite{PhysRevX.2.041014,PhysRevLett.110.133001,10.21468/SciPostPhys.10.3.052}. In that case the standard deviation $\sigma_{\Delta}$ can be decreased to 0.155 MHz (for $T_a=10$ $\mu$K) which satisfies the condition of $\sigma_{\Delta}/\Omega\ll1$ for suppressing the motion-induced dephasing as proposed in \cite{PhysRevApplied.13.024008}. At $T_a=10$ $\mu$K our numerical simulation shows the Doppler dephasing error can be dramatically lowered to a level of $\sim 10^{-5}$ and safely ignored for all cases. Additionally, advanced laser technology that depends on low-noise laser sources can further eliminate phase errors  \cite{PhysRevLett.121.123603}.

In addition to the motion-induced dephasing error, we also investigate the influence of the fluctuation of the laser amplitude on the gate fidelity. Generally speaking, such gate protocols based on optimization require an exact knowledge of the modified pulse shapes so as to be sensitive to the fluctuations of laser amplitudes. This feature shares similarities with adiabatic pulses \cite{PhysRevA.101.062309}. Especially, the modified pulse shapes are crucial for the success of the gate procedure. To study this effect, we assume a constant fluctuation $\delta\Omega$ on the Rabi frequency and keep its waveform unchanged, resulting in $\Omega_{1(2)}(t)\to\Omega_{1(2)}(t)+\delta\Omega$ for the two-pulse scheme and $\Omega(t)\to\Omega(t)+\delta\Omega$ for the one-pulse scheme. 
More precisely the deviation $\delta\Omega$ is defined by $\delta\Omega=0.5\Omega_{1(2)}^{pp}\delta_{\Omega}$, where $\Omega_{1(2)}^{pp}$ is the peak-peak value of the modified(optimized) pulse amplitudes and $\delta_{\Omega}\in[0,0.05]$ is a coefficient for relative deviation. A similar denotation with  $\delta_{\Omega}=\delta\Omega/0.5\Omega^{pp}$ is used for the one-pulse scheme. For each $\delta_{\Omega}$ the numerical result is shown in Fig.6(b1-b2) where we impose a random deviation within the range of $[-\delta\Omega,+\delta\Omega]$ on the optimized pulse amplitudes, and carry out the gate procedure. Every point in the figure is an average result over sufficient gate implementations. From the results in  Fig.6(b1) and (b2) we can find the imperfection of the gate fidelity is only $\sim 10^{-4}$(see the insets of (b1-b2)) which certifies the stronger robustness of our two-pulse protocol to the fluctuation of laser amplitudes. A dramatic increase of the gate imperfection with $\delta_\Omega$ is observed in the one-pulse case as a consequence of a larger peak-peak amplitude $\Omega^{pp}$. With the relative deviation $\delta_\Omega$ up to $5\%$ we show that the imperfection becomes $\sim0.05$($0.005$) for $V_0/2\pi=7.0(1.0)$ MHz.
We expect a long-term stable and precisely-adjustable Rydberg laser system in the experiment, which can deeply reduce this technical error \cite{Arias:17}.


\section{Three-qubit Toffoli gate}

So far we focus on a straightforward implementation of two-qubit CNOT gates. Next, we proceed to describe how to realize a three-qubit Toffoli gate with the same optimization algorithm. Consider three atoms arranged in a line where two outer control atoms constrain the behavior of the central target atom, we assume the nearest neighboring interaction is  $V_0/2\pi=7.0$ MHz, leading to a weak next-nearest neighboring interaction $V_0/2^6=2\pi\times0.11$ MHz
between two control atoms. Its influence is avoidable when one adopts a two-dimension triangular arrangement \cite{PhysRevLett.112.183002}, but for generality, we describe the gate in the line arrangement. Apparently, if any control atom is prepared in the idle state $|1\rangle$, the three-qubit inputs can be directly reduced to the state same as in the two-qubit gate, which are
\begin{equation}
    |10\beta\rangle\Rightarrow|0\beta\rangle,|01\beta\rangle\Rightarrow|0\beta\rangle,|11\beta\rangle\Rightarrow|1\beta\rangle.
\end{equation}
with $\beta\in\{0,1\}$. Discussions for implementing a two-qubit CNOT gate have been given in Sec. III. 
So here we only pay attention to the most different case if the input states are $|000\rangle$ and $|001\rangle$, which means two outer control atoms acquire a same ground-Rydberg coupling by the optical pulses. Intuitively the weak next-nearest neighboring interaction between them would facilitate a simultaneous excitation onto state $|rr\rangle$ by following the ideal transformation of
\begin{equation}
    |00\beta\rangle\to|rr\beta\rangle\to|00\beta\rangle
\end{equation}
prohibiting the excitation of the target atom.
Nevertheless, the realistic three-atom dynamics is very complicated due to the leakage of the target atom onto other unwanted states under the globally-optimized laser pulses. Therefore we use numerical optimization based on same genetic algorithm to construct amplitude and phase-modulated pulses for the three-qubit Toffoli gate.

Utilizing optimization algorithm we numerically explore the realization of a three-qubit Toffoli protocol, in which the gate fidelity is assumed as
\begin{equation}
   \mathcal{F}_3=\frac{1}{8}\text{Tr}\left[\sqrt{\sqrt{U_\text{Toffoli}}\rho_{det}(t=T_g)\sqrt{U_\text{Toffoli}}}\right] 
   \label{Ftoff}
\end{equation}
with the transformation matrix given by
\begin{equation}
    U_{\text{Toffoli}}=\left( 
\begin{array}{cccccccc}
1 & 0 & 0 & 0& 0 & 0 & 0 & 0 \\ 
0 & 1 & 0 & 0& 0 & 0 & 0 & 0  \\ 
0 & 0 & 1 & 0& 0 & 0& 0 & 0  \\   
0 & 0 & 0 & 1& 0 & 0& 0 & 0  \\
0 & 0 & 0 & 0& 1 & 0& 0 & 0  \\
0 & 0 & 0 & 0& 0 & 1& 0 & 0  \\
0 & 0 & 0 & 0& 0 & 0& 0 & 1  \\
0 & 0 & 0 & 0& 0 & 0& 1 & 0  
\end{array}%
\right)  \label{Toffoli} \\ 
\end{equation}
and the eight computational basis states $\{|000\rangle$, $|001\rangle$, $|010\rangle$, $|011\rangle$, $|100\rangle$, $|101\rangle$, $|110\rangle$, $|111\rangle\}$. The density matrix $\rho_{det}$ in Eq.(\ref{Ftoff}) is solved from the master equation with two-body interactions. The optimal pulses have a duration of $T_g=1.2$ $\mu$s. \textcolor{black}{Notice that this three-qubit Toffoli gate is achieved straightforwardly with a pure {\it vdWs} energy shift $V_0/2\pi=7.0$ MHz between two nearest atoms.} No extra local single-qubit operations are required that leads to the total gate time being as short as $1.2$ $\mu$s. Note that in Ref. \cite{PhysRevA.98.042704} a fast three-qubit Toffoli gate was proposed via three-body F\"{o}rster resonance with a multi-pulse sequence of 2.46 $\mu$s-duration.

\begin{figure}
\includegraphics[width=0.5\textwidth]{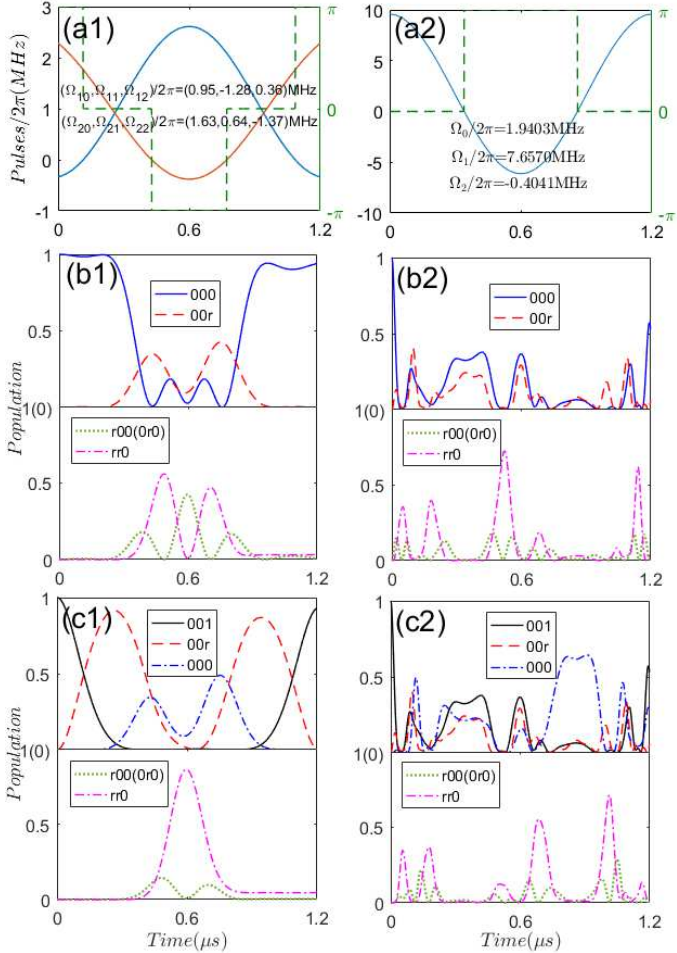} 
\caption{Realization of a three-qubit Toffoli gate. (a1) The optimal pulse amplitudes and the relative phase $\delta\phi$ in the two-pulse scheme. (b1-c1) The corresponding time-dependent population dynamics of the input states $|000\rangle$ and $|001\rangle$. The nearest-neighbor interaction is $V_0/2\pi=7.0$ MHz and the decay rate from Rydberg states is $\gamma=3.0$ kHz. Similarly, (a2-c2) show the results of the one-pulse scheme. State evolutions of $|r00\rangle$ and $|0r0\rangle$ are same and marked by same linetype. Without considering the fluctuation of interactions, the overall gate fidelity is $\mathcal{F}_3 = 0.9738$(two-pulse) and $\mathcal{F}_3=0.9039$(one-pulse).}
\label{three} 
\end{figure}

For the sake of clarity in Fig. \ref{three} we only show the population dynamics when the input states are $|000\rangle$ or $|001\rangle$. First, we observe that when the target atom is initialized in state $|0\rangle$, {\it i.e.} $|000\rangle$, the simultaneous excitation of three atoms caused by $\Omega_1(t)$ will arise a competitive effect. Because two control atoms experience a dynamical swapping of
\begin{equation}
  |rr0\rangle\rightleftarrows\frac{1}{\sqrt{2}}(|0r\rangle+|r0\rangle)\otimes|0\rangle  
\end{equation}
before returning back to $|000\rangle$ [lower panels, Fig. \ref{three}(b1-b2)]. During this period the target atom is also possibly excited to $|r\rangle$ due to the use of a global pulse [upper panels, Fig. \ref{three}(b1-b2)]. If the input state is $|001\rangle$, the competitive effect between the doubly- and singly-excited Rydberg states for two control atoms persists in the dynamics [lower panels, Fig. \ref{three}(c1-c2)]. In addition, we find that state $|001\rangle$ also couples to $|000\rangle$ via a two-photon process, accompanied by a swapping of the target atom between $|001\rangle$ and $|000\rangle$ mediated by $|00r\rangle$ [upper panels, Fig.\ref{three}(c1-c2)].
The central physics here is that the two control qubits, undergoing a simultaneous excitation, can effectively manipulate the behavior of the target atom. The major limitations on the attainable fidelity of a three-qubit Toffoli gate still come from the decay error from the transient Rydberg population, as well as the imperfection in optimization. With intrinsic errors included, the gate fidelity would be $\mathcal{F}_3=0.9738$(or 0.9039) for the two-pulse(one-pulse) scheme.

It is apparent that the optimization algorithm becomes inefficient, especially for the one-pulse case, leading to a relatively poor gate fidelity. That is caused by the presence of fast and irregular population oscillations among various middle states. To our knowledge, most of the evolutionary methods including the genetic algorithm we used, adopt random policies to search for the global maximum. And they perform insufficiently when too many random processes are involved at the same time, {\it e.g.} the one-pulse scheme. While there is no silver bullet, we think the way of reinforcement learning may be an exceptive candidate for solving this problem in which the final fidelity of quantum gate can be highly determined by the choice of a suitable value function  \cite{1998Reinforcement,RL2,RL3,RL4}. More details for this issue can be found in Appendix.\ref{appendix}.





\section{Discussion and Conclusion}


To implement a real one-pulse CNOT gate, one may follow the sketch as illustrated in Fig.\ref{modelthy}(a). By using an AEOM together with an arbitrary waveform generator, the amplitude of the single laser beam can be modulated as requested according to the optimization {\cite{Zhou:16}}. A subsequent AOM can spatially separate the single pulse into the zeroth-order and the first-order diffraction pulses with a hyperfine frequency difference $\omega_{01}\sim $ GHz. 
Ideally, the zeroth-order field serves as pulse 1 that globally excites both atoms, realizing the off-resonant transition from $|0\rangle$ to the intermediate state {\it e.g.} $|5P_{3/2}\rangle$. While the first-order field deviates from the zeroth-order one in space which couples states $|1\rangle$ and $|5P_{3/2}\rangle$ of the target atom, serving as the local pulse 2. Here we assume the coupling strengths between $|0\rangle$ and $|1\rangle$ with the intermediate state are same so as to share a same pulse waveform.
The additional 480 nm laser is left on making transitions of $|5P_{3/2}\rangle\to|70S_{1/2}\rangle$ for both atoms. In fact even if the dipole matrix elements between $|0\rangle$ and $|1\rangle$ with the intermediate state are slightly different, a same effective two-photon Rabi frequency may still be attainable by using an adjustable detuning to the intermediate state \cite{PhysRevA.97.053803}.
In experiment it remains possible to generate same-shaped pulses based on two independent lasers with a fixed frequency difference which requires extra frequency locking technique \cite{doi:10.1063/1.1805272}.
A more detailed study of the implementation of one-pulse CNOT gates, taking into account the practical light-atom coupling mechanism with real energy levels, will be the subject of a future work.


In conclusion, we have presented an experimentally-accessible proposal of the two-qubit CNOT gate and the three-qubit Toffoli gate with weakly-interacting Rydberg atoms using a minimal number of optical pulses. The key lies in a careful optimization of the pulse amplitudes and phases in advance, which leads to the scheme being straightforward for implementation. We adopt a genetic algorithm to perform global optimization which ensures not only a fast gate duration, $\sim 1.0$ $\mu$s for the two-qubit gate($\sim1.2$ $\mu$s for the three-qubit gate), but also a high gate fidelity.
Under practical parameters excluding the fluctuations of interaction, we show that it is possible to create a two-qubit CNOT gate by two(one) optical pulses
with a fidelity of 0.9951(0.9935) when two atoms are separated by 9.76 $\mu$m, and a three-qubit Toffoli gate with a fidelity about 0.9738(0.9039) if the nearest spacing is 7.10 $\mu$m. The more compact scheme with only one pulse achieves a slightly smaller fidelity, because of the imperfection in the optimization algorithm (see Appendix A).

Besides, we systematically investigate various intrinsic and technical errors of the scheme. \textcolor{black}{Compared to the earlier multi-pulse proposals in the strong blockade regime
it is apparent that our entangled gate using simultaneous excitation with fewer pulses works well in the {\it vdWs}-interaction regime, which waives the requirement of fast switching of multiple lasers at different locations and greatly reduces the complexity of gate implementation.} Nevertheless, we also find that our optimization method becomes insufficient while dealing with the most exotic case where the three-qubit Toffoli gate is achieved by one pulse (the fidelity is only 0.9039). Because various ground-Rydberg transitions and interactions make the three-atom dynamics very complicated. In our future work, we will seek more powerful optimization algorithms for the one-pulse Toffoli gate, hopefully achieving a fidelity greater than 0.99.

\begin{acknowledgments}
We are grateful to Shilei Su for help with the numerical simulation, to Jiefei Chen and Jinxian Guo for useful discussions to the experimental design. This work is supported by the National Key Research and Development Program of China under Grant No. 2016YFA0302001;
by the NSFC under Grants No. 12174106, No.11474094,  No.11104076 and No.11654005, by the Science and Technology Commission of Shanghai Municipality under Grant No.18ZR1412800, by the Shanghai Municipal Science and Technology Major Project under Grant No. 2019SHZDZX01, the Shanghai talent program.
\end{acknowledgments}

\appendix

\section{Proof of the {\it vdWs} interactions}

\begin{table*}
\caption{\label{tab:tablex}%
\textcolor{black}{Non-resonant dipole-dipole processes between $|rr\rangle=|70S_{1/2},m_J=1/2;70S_{1/2},m_J=1/2\rangle$ and other nearby Rydberg pairs $|r_{c_j}r_{t_j}\rangle$ with $j\in(1,2,3,4)$, which are $|r_{c_1}r_{t_1}\rangle=|70P_{3/2},m_J=3/2;69P_{3/2},m_J=3/2\rangle$, $|r_{c_2}r_{t_2}\rangle=|70P_{3/2},m_J=3/2;69P_{1/2},m_J=-1/2\rangle$, $|r_{c_3}r_{t_3}\rangle=|69P_{3/2},m_J=3/2;70P_{1/2},m_J=-1/2\rangle$, $|r_{c_4}r_{t_4}\rangle=|70P_{1/2},m_J=-1/2;69P_{1/2},m_J=-1/2\rangle$ obtained from \cite{PhysRevA.77.032723}. The F{\"o}rster energy defects $\delta_{j}$ with respect to state $|r_{c_j}r_{t_j}\rangle$ are $(\delta_1,\delta_2,\delta_3,\delta_4)/2\pi = (0.71,1.01,0.99,1.29) $ GHz. $B_{j}=C_3^{(j)}/r_0^3$ represents the coupling strength with the dispersion coefficients given by $C_3^{(j)}/2\pi=(7.94,6.37,6.59,5.28)$ GHz$\cdot\mu$m$^3$. $\mathcal{E}_j$ stands for the population leakage due to single coupled pair between $|rr\rangle\rightleftarrows|r_{c_j}r_{t_j}\rangle$, and $\bar{\mathcal{E}}$ gives the average population leakage from state $|rr\rangle$ when all channels $|r_{c_j}r_{t_j}\rangle$ are considered.
}}
\begin{ruledtabular}
\setlength{\tabcolsep}{0mm}{
\begin{tabular}{cccccccccc}
$r_{0}$($\mu m$)&$B_{1}/2\pi$(MHz)&$\mathcal{E}_{1}$&$B_{2}/2\pi$(MHz)&$\mathcal{E}_{2}$&$B_{3}/2\pi$(MHz)&$\mathcal{E}_{3}$&$B_{4}/2\pi$(MHz)&$\mathcal{E}_{4}$&$\bar{\mathcal{E}}$\\
\hline
4.89&67.90&$1.7\times10^{-2}$&54.45&$5.7\times10^{-3}$&56.33&$6.4\times10^{-3}$&45.17&$2.5\times10^{-3}$&$2.5\times10^{-2}$\\
\hline
7.10&22.18&$1.9\times10^{-3}$&17.79&$6.1\times10^{-4}$&18.40&$6.9\times10^{-4}$&14.76&$2.6\times10^{-4}$&$2.9\times10^{-3}$\\
\hline
9.76&8.54&$2.8\times10^{-4}$&6.85&$9.1\times10^{-5}$&7.08&$1.0\times10^{-4}$&5.68&$3.9\times10^{-5}$&$4.3\times10^{-4}$\\
\end{tabular}
}
\end{ruledtabular}
\end{table*}

To search for an appropriate interatomic distance $r_0$ that can be used for a weak {\it vdWs} interaction between two distant atoms, we numerically simulate the population dynamics of state $|rr\rangle$ due to the non-resonant interaction with pairs of Rydberg states $|r_{c_j}r_{t_j}\rangle$.
Here we apply four dominant states, yet in principle we have to sum up sufficient adjacent states at the same time which leads to a second-order level shift of $|rr\rangle$ represented by the coefficient $C_6\approx \sum_j (C_3^{(j)})^2/\delta_j$ \cite{PhysRevA.96.042306}. 
Because this non-resonant interaction is only relevant to the population of state $|rr\rangle$ we choose it as the initial state in the calculation.

\begin{figure}
\centering
\includegraphics[width=0.4\textwidth]{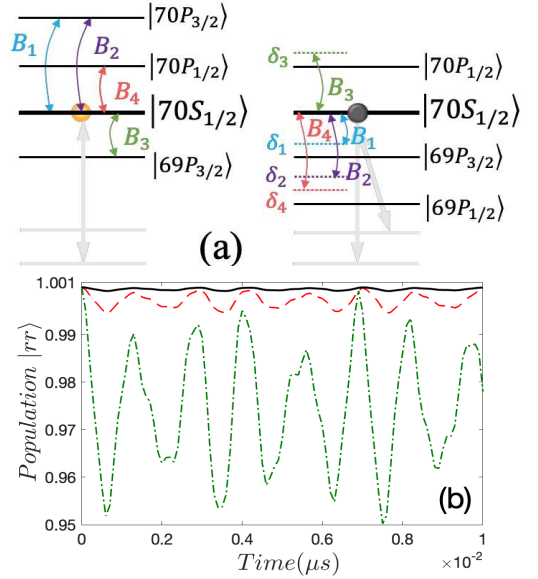} 
\caption{(a) Several dominant leakage channels of the two-atom Rydberg states for the detuned dipole-dipole interactions where $|rr\rangle$=$|70S_{1/2},m_J=1/2;70S_{1/2},m_J=1/2\rangle$ is initially occupied. $B_j$ and $\delta_j$($j=1,2,3,4$) represent the coupling strength and the F\"{o}rster energy defect between $|rr\rangle$ and $|r_{c_jr_{t_j}}\rangle$. (b) Population dynamics of state $|rr\rangle$ taking account into four nearby two-atom Rydberg states. Different central distances $r_{0}=4.89$ $\mu m$(green dash-dotted), $7.10$ $\mu m$(red dashed) and $9.76$ $\mu m$(black-solid) are used. Detailed parameters can be found in Table II. }
\label{dipole} 
\end{figure}

We consider four dominant leakage channels as shown in Fig. \ref{dipole}(a) where the coupling strengths $B_j$ and the corresponding F{\"o}rster defects $\delta_j$ are given in Table II using the ARC open source library {\cite{SIBALIC2017319}}. In Fig. \ref{dipole}(b) we show the time-dependent population dynamics of state $|rr\rangle$ with four leakage channels under different two-atom distances $r_0=(4.89,7.10,9.76)$ $\mu$m. It is clearly noted that the average leakage of population is around $\bar{\mathcal{E}}\sim 0.025$ for $r_0=4.89$ $\mu$m which will cause a big effect on the optimal gate fidelity which is typically larger than $>99\%$ if the fluctuation of interactions is excluded. In this case it is inappropriate to assume a pure {\it vdWs} interaction for state $|rr\rangle$. However, if the two-atom distance is increased to be $r_0=7.10$ $\mu$m or $9.76$ $\mu$m which are larger than Rydberg blockade distance, the population missing from state $|rr\rangle$ acquires a dramatic reduction. Depending on a full calculation including all(four) leakage channels we show the imperfection due to non-resonant dipole-dipole interactions can be at the level of $\bar{\mathcal{E}}\sim 2.9\times 10^{-3}$ for $r_0=7.10$ $\mu$m and $\bar{\mathcal{E}}\sim4.3\times 10^{-4}$ for $r_0=9.76$ $\mu$m, which are smaller than the intrinsic error during the gate execution. In the article we would assume two well-separated atoms with a distance of $7.10$ $\mu$m and 9.76 $\mu$m,
arising a {\it vdWs} interaction of 
 $V_0=C_6/r_0^6=7.0$ MHz and $1.0$ MHz as used in our calculation.

\section{Numerical optimization algorithm}
\label{appendix}

\begin{figure}
\includegraphics[width=0.5\textwidth]{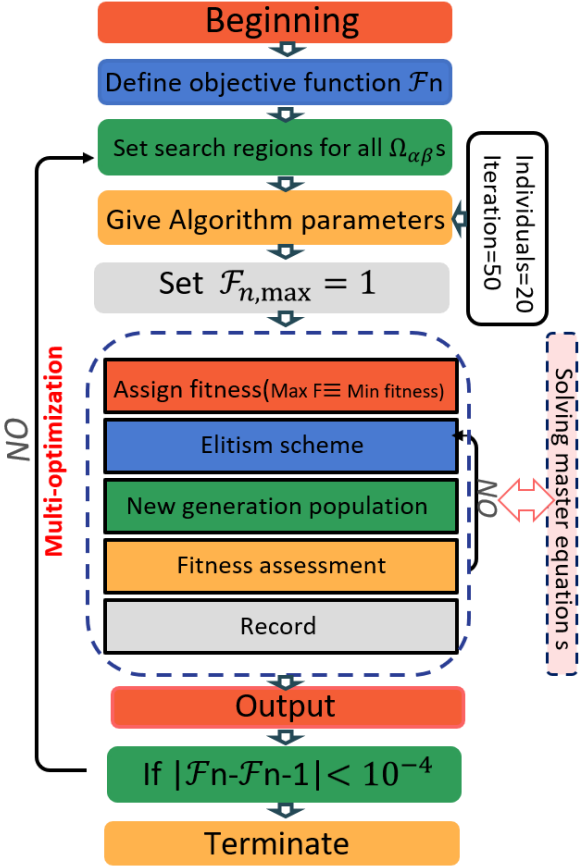} 
\caption{Graphic diagram for the genetic optimal algorithm where the dashed box shows the core of algorithm that connects with the physical model. }
\label{apped} 
\end{figure}

{\it Genetic algorithm description.}  
Utilizing numerical optimization procedures we can obtain a set of  optimal pulse parameters which approximately implements the gate with a high performance. We adopt genetic algorithm(GA) in this article.
GA is a kind of random search algorithm inspired by Darwin’s theory of natural selection and evolution. It can realize heuristic search for a complex search space via simplified genetic process, which usually consists of the population initialization, fitness assessment, selection, crossover and mutation. 
The aim of GA's usage in our study is to find out a set of parameters in the given population that can make one of the individuals in the population has the best fitness. The fitness is determined according to the objective function. 
GA is only determined by the fitness function that has no limit on the domain of definition, which robustly extends its generality. GA starts optimization from a series of temporary results and iterates them simultaneously which allows it to avoid local maximum and realize parallel computation easily. Thus, GA offers a global optimization within the given search region for all parameters, without the need for parameter initialization. \textcolor{black}{However, Rudolph has shown that the traditional canonical GA, which only contains crossover, mutation and selection, can hardly realize a global optimization because the crossover and mutation operators may destroy the high-fitness scheme of an individual \cite{Rudolph}.} Thus, Ref. \cite{DeJong} proposed an elitism scheme to overcome this which is verified to get a global convergence. \textcolor{black}{In our work, we adopt the method of Ref.\cite{DeJong} which can always preserve the best individual without crossover or mutation to the next generation.}


    According to the procedure diagram as shown in Fig.\ref{apped} we start by assuming an objective function {\it i.e.} the targeted gate fidelity $\mathcal{F}_n$ with $n$ denoting the number of optimizations. We ignore the subscripts 2 or 3 as defined in Eq.(9) and (15) for the qubit number. Then we give a reasonable initial search region
    for all pulse parameters $\Omega_{\alpha\beta}s$. For example, in the two-pulse scheme the set of parameters is 
    $(\Omega_{10},\Omega_{11},\Omega_{12})$ and $(\Omega_{20},\Omega_{21},\Omega_{22})$. The number of loop iterations in the algorithm is set to be 20$\times$50. Here 20 is for the population size and 50 means the evolutionary generation number. The maximum of objective function is $\mathcal{F}_{n,max}=1$. By solving the master equation (7) with random pulse parameters $\Omega_{\alpha\beta}$s which are obtained from \textcolor{black}{the initial search region}, we start to perform 20$\times$50 iterations by GA and get a gate fidelity of $\mathcal{F}_1$ after the first optimization. To improve the precision, we continuously perform the $(n)$th optimization where the new search region is given depending on the optimized pulse parameters obtained after the $(n-1)$th optimization. $n$ is the number of optimizations. Finally, if the criterion $\delta\mathcal{F}=|\mathcal{F}_n-\mathcal{F}_{n-1}|<10^{-4}$ is met the iteration is terminated; otherwise, we restart the $(n+1)$th optimization until this criterion is reached. All numerical results presented in the paper are obtained by multi-optimization with same precision.

 \begin{table*}
\caption{\label{tab:table3}%
\textcolor{black}{For constructing a two-qubit CNOT gate with one pulse or two pulses, the optimized pulse parameters and the corresponding gate fidelities are presented with different {\it vdWs} interactions. Here $C_6/2\pi = 863$ GHz$\cdot\mu m^6$, $\gamma =3.0 $ kHz, $\mathcal{F}_{2,one(two)}$ represents the two-qubit gate fidelity in the one(two)-pulse scheme, and all frequency parameters are in unit of MHz. Numerical plot of these results is shown in Fig.5(b).  }}
\begin{ruledtabular}
\setlength{\tabcolsep}{0mm}{
\begin{tabular}{ccccccc}
$V_{0}/2\pi$&$r_{0}$($\mu m$)&($\Omega_{0}$,$\Omega_{1}$,$\Omega_{2})/2\pi$&$\mathcal{F}_{2,one}$&($\Omega_{10}$,$\Omega_{11}$,$\Omega_{12})/2\pi$&$(\Omega_{20}$,$\Omega_{21}$,$\Omega_{22})/2\pi$&$\mathcal{F}_{2,two}$\\
\hline
1.0& 9.76&(-5.8283,-5.5942,1.3558)& 0.9935&(-0.9549,-1.9544,-0.0631)&(-2.8310, -3.6488,0.0074)&0.9951\\
2.0& 8.69&(-8.0524, -2.3823, -6.2005) & 0.9938 & (-3.2197, -1.6527, -4.3272)&(-5.5704, 0.5088, -5.5704)&0.9897 \\
3.0& 8.13&(-0.8549,-15.0568,4.5848)& 0.9783&(-6.4985,7.2516,0.2753)&(-0.0947, -3.0809, 0.5679)&0.9752\\
4.0& 7.74&(-2.6977,-15.8330,7.5137)& 0.9810&(1.6474,-2.2875,1.2089)&(0.5896, 1.7279,0.0084)&0.9723\\
5.0& 7.46 &(6.1341,-11.1408,9.1814)& 0.9840&(-1.4798,-1.0744, -4.0572)&(-1.2613,4.4488,5.7295)&0.9804\\
6.0& 7.24&(-0.5748, -15.1009, 8.7235) & 0.9899 & (-1.4115, -2.2769, -2.3879)&(0.5608, -1.6526, -1.5915)&0.9834 \\
7.0& 7.10&(-0.7619, -15.7833, 8.9923) & 0.9923 & (-3.9621, -0.7858, 1.5915)&(1.0942, -1.9068, -2.2182)&0.9921 \\
\end{tabular}
}
\end{ruledtabular}
\end{table*}

{\it Failure of the genetic algorithm.} Furthermore, we show some details about why GA performed badly when implementing the three-qubit Toffoli gate with one pulse, and why we expect reinforcement learning(RL) as a candidate for this problem.
To our knowledge, evolutionary methods including GA, apply some static policies which interact over a period of time or one agent's life with a separate instance of the environment \cite{1998Reinforcement}. In other words it means that the son generation could only perceive and interact with what its parents left. Thus, those evolutionary methods are effective only if the space of policy in optimization is small enough or could be cut apart into a set of small spaces. 
Unfortunately, the \textcolor{black}{search} space is hard to be separated in our system. On the other hand, those methods with more randomness are always seeking for various policies or ``paths", in order to refine the results. 
In general, a traditional evolutionary method is difficult to obtain sufficient information
if the process of evolution is highly random in some specific tasks. For example, in the case of three-qubit Toffoli gate with one pulse 
as described in Fig.\ref{three}(a2-c2), the poor performance of states $|000\rangle$ and $|001\rangle$ arises from the fact that GA can not acquire sufficient evolutionary information from the fast- and irregular-oscillating dynamics.

In the future we expect the RL method can become a candidate for optimizing parameters in constructing a high-fidelity Toffoli gate with one pulse. Compared to evolutionary methods, RL includes dynamical policies with the help of value functions. As a consequence, RL can learn the feature of the environment by interacting with it, and can use more information to find out a better outcome than evolutionary methods. The targeted fidelity of quantum gates is determined by the 
selection of an appropriate value function. Therefore in RL by using a proper reward policy it is desirable to improve the performance of a one-pulse Toffoli gate protocol \cite{lee2021efficient}. Recently, we propose a new deep learning method with a hybrid network for solving the high-frequency oscillating population dynamics in a double-well potential, paving one-step closer to this target \cite{li2021revisiting}.
 
\section{Optimized pulse parameters under different central interactions}

As a supplement to Fig.5(b),
Table \ref{tab:table3} presents the optimized pulse parameters as well as the simulated gate fidelity within an adjustable {\it vdWs} interaction regime $V_0/2\pi\in[1.0,7.0]$ MHz. No fluctuated interaction is considered.

\bibliography{apssamp}

\end{document}